\documentclass[12pt]{iopart}

\usepackage{graphics}
\usepackage{graphicx}
\usepackage{epsfig}

\begin{document}

\title[Black Hole Backgrounds]{Pulsar Timing Array Analysis for Black Hole Backgrounds}

\author{Neil J. Cornish$^1$ and A. Sesana$^2$}

\address{$^1$ Department of Physics, Montana State University, Bozeman,
MT 59717, USA}

\address{$^2$ Max Planck Institut fŸr Gravitationsphysik (Albert-Einstein-Institut), D-14476 Potsdam, Germany}

\begin{abstract}
An astrophysical population of supermassive black hole binaries is thought to be the strongest source of gravitational waves in the frequency range covered by
Pulsar Timing Arrays (PTAs). A potential cause for concern is that the standard cross-correlation method used in PTA data analysis assumes that the signals are
isotropically distributed and Gaussian random, while the signals from a black hole population are likely to be anisotropic and deterministic. Here we argue
that while the conventional analysis is not optimal, it is not hopeless either, as the standard Hellings-Downs correlation curve turns out to hold for point
sources, and the small effective number of signal samples blurs the distinction between Gaussian and deterministic signals. Possible improvements to the
standard cross-correlation analysis that account for the anisotropy of the signal are discussed.
\end{abstract}



\maketitle

\section{Introduction}

The most promising source of signals in the frequency range covered by Pulsar Timing Arrays (PTAs) is from a population of supermassive black hole
binaries, dominated by systems with masses in the range $3 \times 10^7 M_\odot \rightarrow 3 \times 10^9 M_\odot$, and times to merger in the
range $10^3 \, {\rm years} \rightarrow 10^5 \, {\rm years}$. It had been assumed that the number density of sources as a function of frequency, $dN/df$, would
be sufficiently large that the central limit theorem would come into play, and that the combined signal would be Gaussian distributed and isotropic. However,
recent studies based on more realistic population synthesis models have shown that the signal is likely to be dominated by a small number of relatively
nearby sources~\cite{Sesana:2008mz, Sesana:2008xk, Sesana:2010mx, Kocsis:2010xa}, and as a result, will be non-Gaussian and anisotropic~\cite{CornishSesana, Ravi:2012bz}.
This is a concern since the standard analysis techniques~ \cite{Hellings:1983fr,Jenet:2005pv,vanHaasteren:2008yh} are based on the assumption
that the signal is isotropic and Gaussian.

 Here we show that, while the standard approach may not be optimal, it is able to detect the signals from isolated black holes, and by extension, populations of
 black holes no matter how sparse. What makes this possible is the rather surprising result that the Hellings-Downs correlation curve~\cite{Hellings:1983fr}, which
 was originally derived for un-polarized, isotropic backgrounds, continues to be valid for polarized point sources! Like many results that are surprising initially,
 after a little thought this result starts to make sense (it is basically a reflection of the quadrupole nature of the signal), and very soon the result becomes obvious, and
 a soon after that, something everyone knew already. 
 
 While the standard cross-correlation analysis technique can be used to detect the signals from a sparse black hole background, it will not be optimal. We consider
 a variety of alternative analysis techniques that may be more effective, and suggest a new cross-correlation technique that accounts for the anisotropy of the signal.

\section{Detector Response}

The timing residuals for a Pulsar located in the $\hat{n} \rightarrow (\theta_p,\phi_p)$ direction,  induced by a plane gravitational wave from a source in
the $(\theta,\phi)$ direction, can be expressed as
\begin{equation}
r = \frac{1}{2} \left( R_+(\cos2\psi F^+ - \sin2\psi F^\times)+R_\times(\sin2\psi F^+ + \cos 2 \psi F^\times) \right) \, ,
\end{equation}
where $\psi$ is the polarization angle of the wave relative to the frame defined by the basis vectors $\hat{u},\hat{v}$ that span the
plane perpendicular to the propagation direction $\hat{k}$, where
\begin{eqnarray}
&& \hat{k} = -(\sin\theta \cos\phi \, \hat{x} +\sin\theta\sin\phi \, \hat{y} +\cos\theta\, \hat{z}) \, , \nonumber \\
&& \hat{u} = \cos\theta \cos\phi \, \hat{x} +\cos\theta\sin\phi \, \hat{y} - \sin\theta\, \hat{z}\, , \nonumber \\
&& \hat{v} = \sin\phi \, \hat{x} -\cos\phi \, \hat{y} \, .
\end{eqnarray}
The antenna bean pattern functions have the form
\begin{eqnarray}
&&F^+ = \frac{(\hat{u}\cdot\hat{n})^2 - (\hat{v}\cdot\hat{n})^2}{1+\hat{k}\cdot\hat{n}} \nonumber \\
&&F^\times = \frac{2(\hat{u}\cdot\hat{n})(\hat{v}\cdot\hat{n})}{1+\hat{k}\cdot\hat{n}}  \, .
\end{eqnarray}
The terms $R_{+,\times}$ are expressed in terms of the anti-derivative, $H_{+,\times}$ of the gravitational wave strain $h_{+,\times}$:
\begin{equation}
R_{+,\times} = H_{+,\times}(t)-H_{+,\times}(t-L(1+\hat{k}\cdot\hat{n})) \, ,
\end{equation}
where $L$ is the distance to the Pulsar from Earth. The two terms in the above equation are referred to as the ``Earth term'' and the ``Pulsar term'', respectively.
For nearby sources the plane wave approximation may need to be augmented by the leading order spherical wavefront corrections of order $L/D$, where $D$ is the distance to the source:
\begin{equation}
R_{+,\times} = H_{+,\times}(t)-H_{+,\times}(t-L(1+\hat{k}\cdot\hat{n})+\frac{L^2}{D}(1-(\hat{k}\cdot\hat{n})^2)) \, .
\end{equation}
The antenna patterns can be re-written in the alternative, simpler form
\begin{eqnarray}
&&F^+ =(1+\cos\beta)\cos(2\alpha) \nonumber \\
&&F^\times =  (1+\cos\beta)\sin(2\alpha)\, ,
\end{eqnarray}
where $\beta= \arccos(-\hat{k}\cdot n)$ is the angle between the source and the Pulsar, and $\alpha=\arctan((\hat{v}\cdot\hat{n})/(\hat{u}\cdot\hat{n}))$
is the angle the Pulsar direction makes relative to the $\hat{u},\hat{v}$ polarization frame. The timing residuals then take the form
\begin{equation}
r = \frac{1}{2} \left( R_+ \cos(2\psi+2\alpha) +R_\times\sin(2\psi+2\alpha) \right)(1+\cos\beta) \, .
\end{equation}

\section{Correlation Analysis}
The cross-correlation of the timing residuals from two Pulsars can be written as
\begin{eqnarray}\label{cross}
\langle r_i r_j \rangle &=& \frac{1}{4} (\langle R_+^2 \rangle  \cos(2\psi+2\alpha_i)\cos(2\psi+2\alpha_j) \nonumber \\
&+& \langle R_\times^2 \rangle  \sin(2\psi+2\alpha_i)\sin(2\psi+2\alpha_j) )(1+\cos\beta_i)(1+\cos\beta_j),
\end{eqnarray}
where the angle brackets denote the inner product
\begin{equation}
\langle h_1  h_2\rangle = \int dt_1 \int dt_2 \; h_1(t_1) K(t_1, t_2) h_2(t_2) \, .
\end{equation}
For stationary signals, the convolution kernel is a function of the lag $\vert t_1-t_2 \vert$, and the inner product can be re-wrriten in the Fourier domain
in the familiar form
\begin{equation}
\langle h_1  h_2\rangle = \int_0^\infty \frac{ 2( \tilde{h}_1(f) \tilde{h}^*_2(f) +\tilde{h}^*_1(f) \tilde{h}_2(f))}{S(f)} \, df \, .
\end{equation}
In (\ref{cross}) it has been assumed that $\langle R_+ R_\times\rangle = 0$, which holds for cosmological stochastic backgrounds and binary systems.
For isotropic gravitational wave backgrounds it makes sense to average the cross correlation over the sky: 
\begin{equation}\label{skyav}
\frac{1}{4 \pi} \int \langle r_i r_j \rangle\,  d\Omega = \langle H^2 \rangle \, \alpha_{ij}\, ,
\end{equation}
where $ \langle H^2 \rangle =  \langle R_+^2 + R_\times^2 \rangle$, and the Hellings-Downs correlation curve $\alpha_{ij}=\alpha(\theta_{ij})$
is given as a function of of the angle $\theta_{ij}=\mu$ between the Pulsars:
\begin{equation}
\alpha(\mu) = \frac{1-\cos\mu }{2} \ln\left( \frac{1-\cos \mu }{2} \right) - \frac{1-\cos \mu }{12}+\frac{1}{3}\left( 1 + \delta(\mu)\right) \, ,
\end{equation}
The delta function - defined such that $\delta(0)=1$, and is otherwise zero - comes from the Pulsar term, which averages to a non-zero value in the auto-correlation.

\begin{figure}[!ht]
  \centering
  \begin{tabular}{cc}
  \includegraphics[width=75mm]{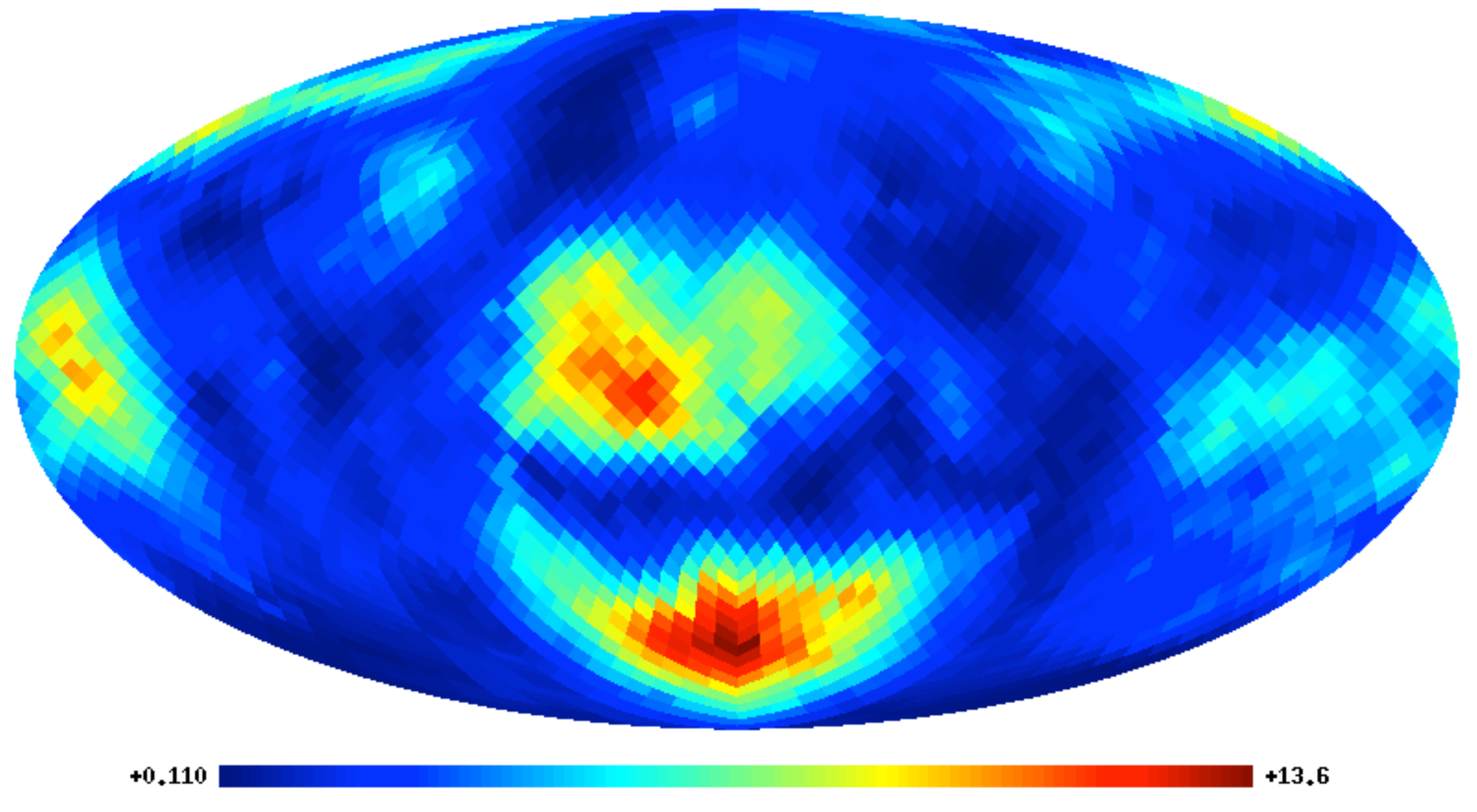}  &
    \includegraphics[width=75mm]{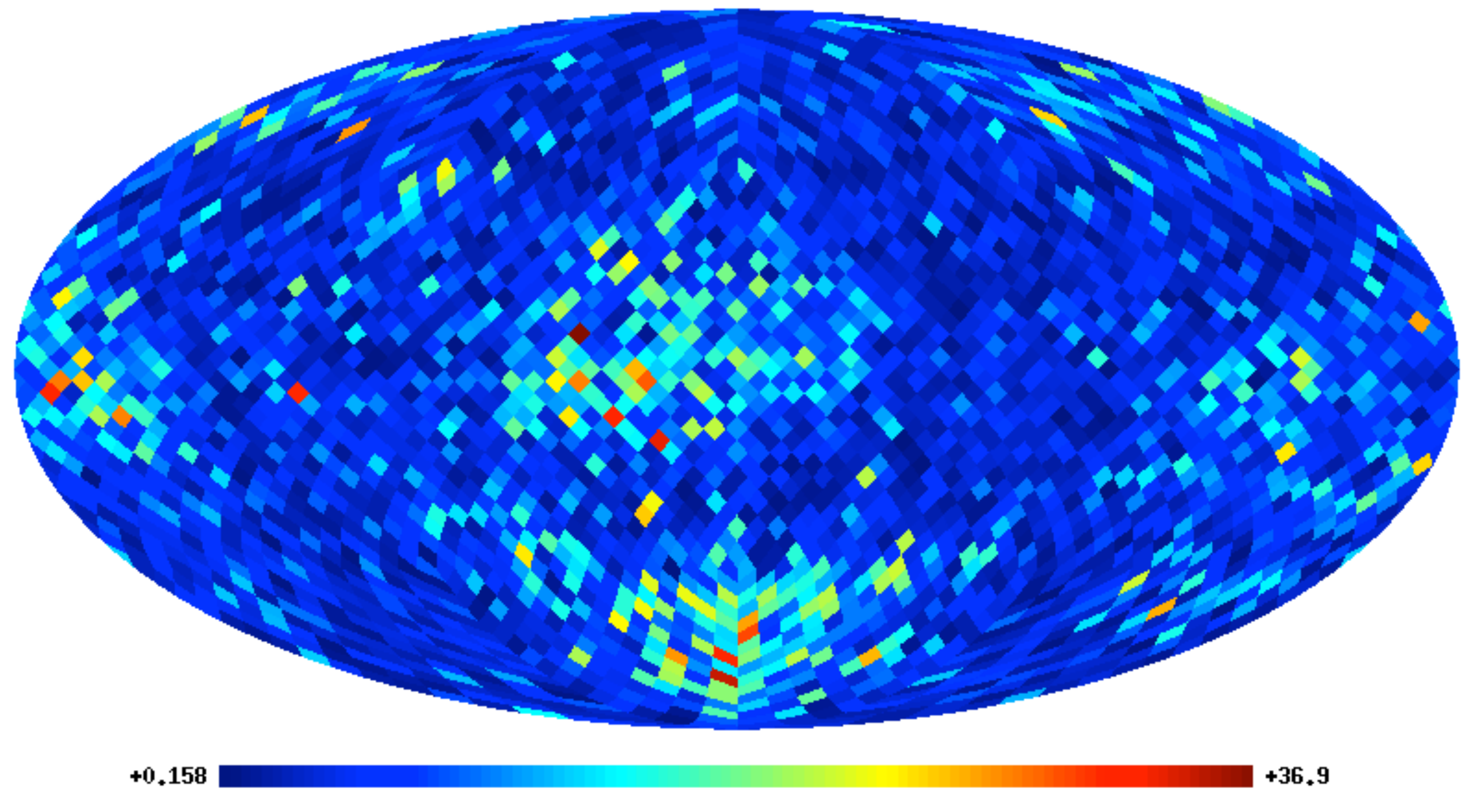} \\
    \end{tabular}
  \caption{The auto-correlated signal power $\langle r^2(\theta_p,\phi_p) \rangle$ for a single realization of a black hole binary population model. The left panel
  has the ``Pulsar terms'' turned off, while the right panel shows the full response. The Pulsar term adds noise, effectively multiplying the Earth-term sky map by
  $2(1-\cos(\delta))$, where $\delta$ is a random phase. Generating the full response at higher angular resolution and then applying a Gaussian smoothing yields a sky map
  nearly identical to the map with the Pulsar term turned off. Either way, the anisotropy of the signal is clear, with Pulsars in certain sky directions receiving significantly
  larger signal power than others.}\label{fig:sky}
\end{figure}

For anisotropic signals, such as those produced by a single black hole binary, sky averaging is not justified, and the correlation $\langle r_i r_j \rangle$ will
depend on the sky location of the source $(\theta,\phi)$, and the orbital orientation given in terms of the inclination and polarization
angles $(\iota,\psi)$. It had been assumed that an astrophysical population of binaries would combine to yield an isotropic, stochastic background, but
this turns out not be be the case. Instead the combined signal is dominated by a handful of nearby, bright sources, and as shown in Figure 1, the resulting
background is highly anisotropic. The BH population model used to generate Figure 1 was derived by extracting catalogues of merging massive galaxies from the
Bertone et al.~\cite{Bertone:2007sj} semianalytic model built on top of the Millennium Run~\cite{Springel:2005nw}. Galaxies were then populated with super massive black holes
correlating with the bulge velocity dispersion as given by Gultekin et al.~\cite{Gultekin:2009qn}. The black holes accrete gas prior to final coalescence and all binaries are assumed
to be circular and driven by GW emission only in the frequency band relevant to PTA. All the steps of the procedure followed to construct the population are given in
Sesana et al.~\cite{Sesana:2008xk}. The anisotropy seen in Figure 1 is even more pronounced if the signal is broken out by frequency bins, where a single source often
dominates in a particular bin. The question then becomes, what is the best technique to detect such a signal, given that it is neither isotropic nor Gaussian?
These assumptions underpin the standard analysis techniques in both the Frequentist and Bayesian implementations. The Frequentist approach is based on the
matched filter detection statistic~\cite{Jenet:2005pv}
\begin{equation}\label{rho}
\rho = \sum_i \sum_{j\neq i} \langle r_i r_j \rangle \, \alpha(\theta_{ij}) \, .
\end{equation}
This statistic is often shifted to have zero mean and scaled to have variance $1/N_{\rm pairs}$, where $N_{\rm pairs}=N(N-1)$ are the number of
pulsar pairs. The key idea is that the pairwise correlations are summed together after being multiplied by the expected correlation function,
which acts like a matched filter. In the Bayesian approach, the correlation function enters into the definition of the multi-variate Gaussian
likelihood function~\cite{Finn:1998vp,vanHaasteren:2008yh},
\begin{equation}\label{like}
p = A \exp\left(-\sum_{i,j} \int ( \tilde{r}_i \tilde{r}^*_j  + \tilde{r}^*_i \tilde{r}_j )  C_{ij}^{-1} df - \frac{1}{2}\int  \ln({\rm det}C) df \right) \, ,
\end{equation}
where $A$ is an overall normalization constant and
\begin{equation}
C_{ij}(f) = S_H(f) \alpha_{ij} + S_{n_i}(f) \delta_{ij} \, .
\end{equation}
Here $S_H(f)$ is the power spectral density of the signal and $S_{n_i}(f)$ is the power spectral density of the noise in the $i^{\rm th}$ Pulsar. In the
weak-signal limit, $S_{n_i} \gg S_H$, the likelihood (\ref{like}) can be approximated as $p = A' \exp(\rho /2)$, drawing out the close connection between the
two approaches.

\section{Isolated Black Hole Binaries}
Before discussing alternative analysis techniques that may be better suited to detecting anisotropic, non-Gaussian signals, it is interesting to consider
how the standard analysis might perform at detecting the signal from an isolated Black Hole binary. To set the stage, let us consider the correlations produced
in a Pulsar Timing Array with 100 randomly distributed Pulsars by (i)  a single black hole binary;  and (ii) an isotropic background. To make the comparison
equitable, the isotropic signal was restricted to a single frequency bin. In Figure 2 the correlations are shown both with and without the Pulsar term, and
un-binned and binned in the angular separation between the Pulsars. The signal strength in each case was scaled to give unit correlation at zero degree
separation. The results in both cases are very similar. The un-binned correlations show significant scatter, while the binned correlations follow the Hellings-Downs
correlation curve. At first sight it may seem strange that an isolated black hole binary produces a correlation pattern that is identical to that produced by
an isotropic background, but on reflection the result is not surprising. The Hellings-Downs curve is simply a consequence of the quadrupolar nature of
gravitational waves. In binning the correlations as a function of the Pulsar angular separation we are replacing the sky average (\ref{skyav}) by
an average over the Pulsar locations, which in the limit of a large number of Pulsar pairs goes over to the integral
\begin{equation}\label{pulav}
\gamma(\mu) = \frac{1}{(4 \pi)^2} \int \langle r_i r_j \rangle\, \delta(\cos\mu - \hat{n}_i \cdot \hat{n}_j) \,  d\Omega_i d\Omega_j \, .
\end{equation}
The Dirac-delta function can be taken care of by adopting a coordinate system where the $j$-Pulsar has coordinates
\begin{eqnarray}
&& x_j = \cos\phi_i(\cos\theta_i \sin\mu \cos\lambda+\sin\theta_i \cos\mu)-\sin\phi_i \sin\mu \sin\lambda \nonumber \\
&& y_j = \sin\phi_i(\cos\theta_i \sin\mu \cos\lambda+\sin\theta_i \cos\mu)+\cos\phi_i \sin\mu \sin\lambda \nonumber \\
&& z_j = \cos\theta_i\cos\mu -\sin\theta_i\sin\mu \cos\lambda\, ,
\end{eqnarray}
which ensure that $ \hat{n}_i \cdot \hat{n}_j = \cos\mu$. Completing the integration over $\lambda, \phi_i,\theta_i$ yields
\begin{equation}
\gamma(\mu) =  \langle H^2 \rangle \, \alpha(\mu) \, .
\end{equation}
Thus, a single black hole produces an identical angular correlation pattern as an isotropic stochastic background. Note that the final
result is independent of the black hole orientation or sky position. Again, this is not surprising since we have integrated the Pulsar locations over the
celestial sphere, which is equivalent to actively rotating the black hole reference frame while holding the Pulsars locations fixed.
\begin{figure}[!ht]
  \centering
  \begin{tabular}{cc}
  \includegraphics[width=75mm]{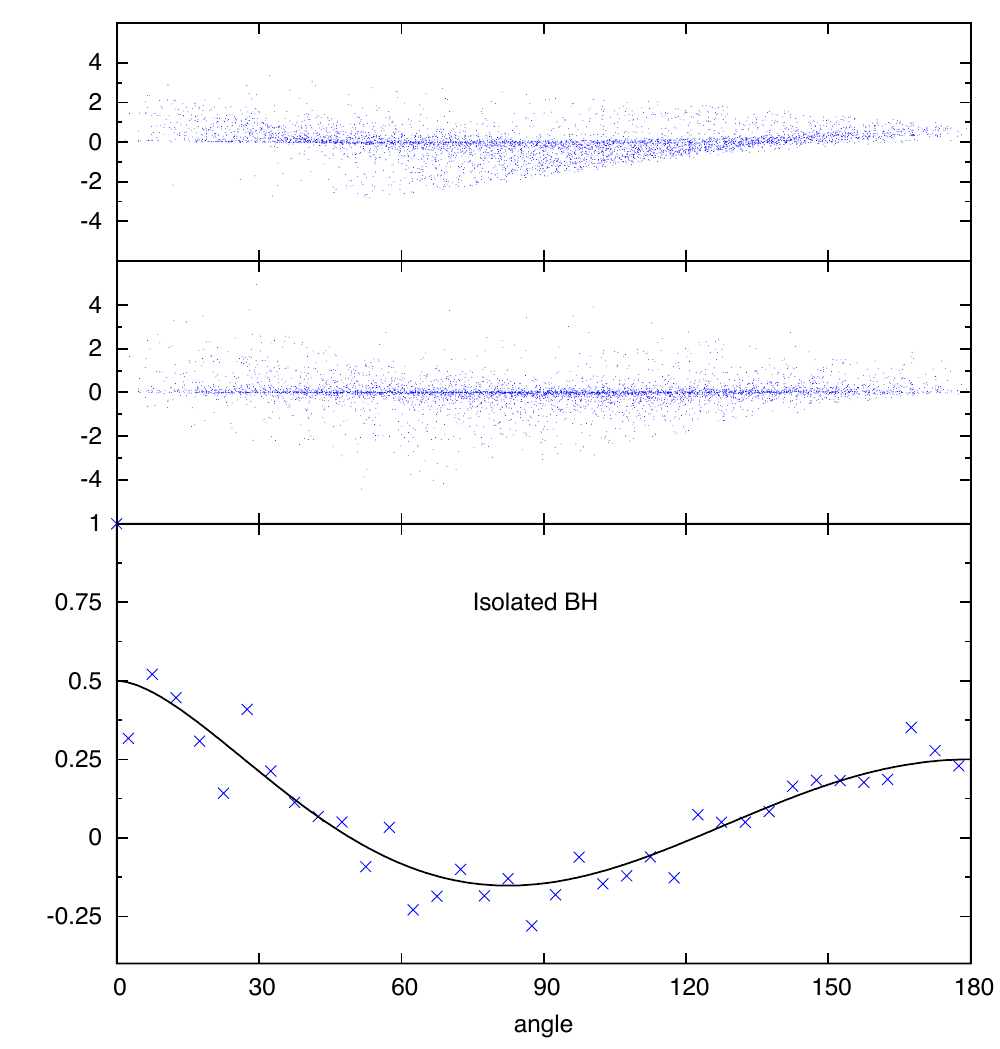}  &
    \includegraphics[width=75mm]{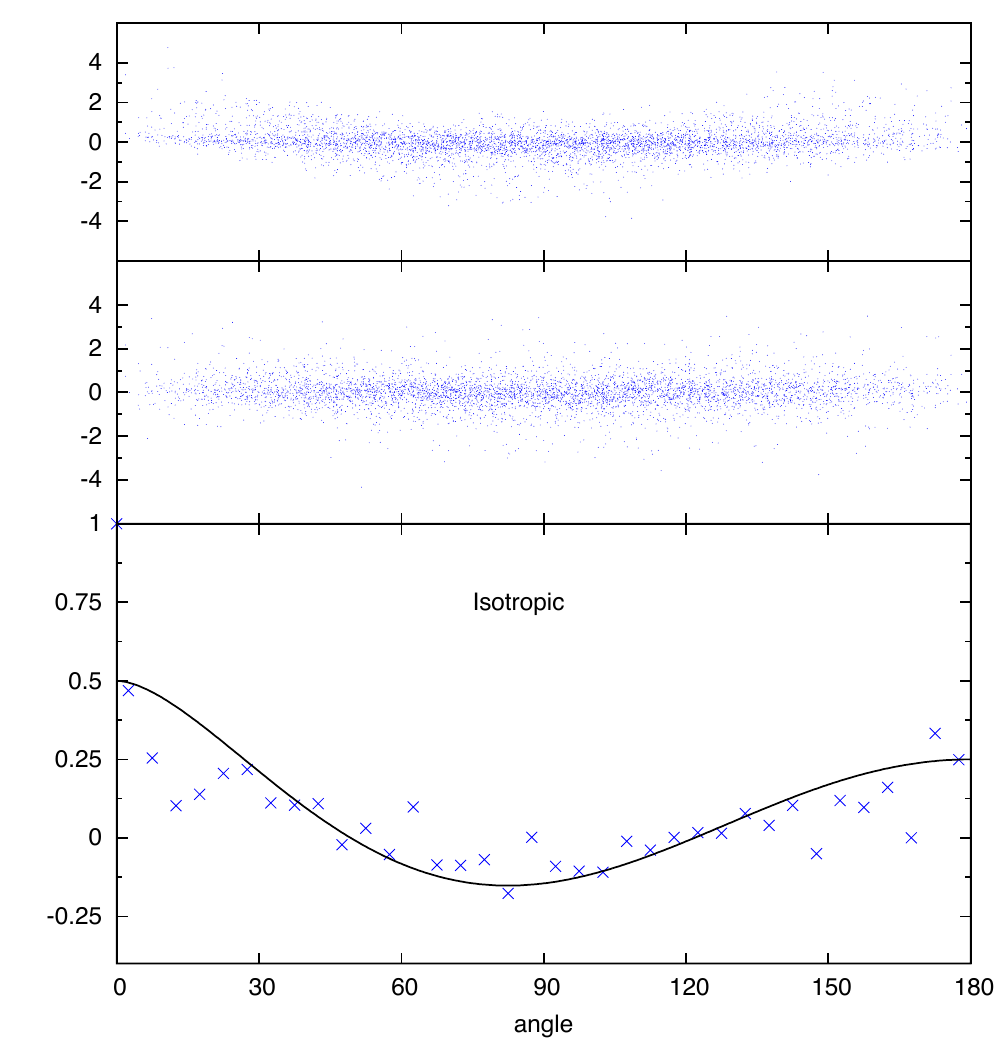} \\
    \end{tabular}
  \caption{Simulated noise free correlation curves for (i) an isolated BH binary (ii) an isotropic background restricted to a single frequency bin.
  The top panels show the correlations as a function of the angle between the Pulsar pairs without the ``Pulsar term''; the middle panels show
  the full correlation;  while the lower panels show the full correlations averaged into 5 degree bins.}\label{fig:comp}
\end{figure}

\section{Astrophysical Black Hole Populations}

In Ref.~\cite{Ravi:2012bz} the applicability of the standard analysis techniques based on (\ref{rho}) and (\ref{like}) for detecting the signals from an astrophysical
population of black holes was discussed. There the focus was on the non-Gaussianity of the signal, rather than the anisotropy. It was noted that
the correlations between Pulsars followed the Hellings-Downs correlation curve upon averaging over $\sim 100$ realizations (unsurprising given
that the averaging restores isotropy), but this result has little practical relevance given that we only get to see a single realization. On the other hand,
the fact that a {\em single} black hole binary yields the Hellings-Downs curve means that the standard analysis techniques will be effective (though
not necessarily optimal) at detecting the signal from a population of black holes. And while it is possible to theoretically establish the non-Gaussianity of the signal
using hundreds of realizations of the population catalogs, it will be difficult to establish in practice with the handful of frequency bins
available for the analysis. Indeed, the departure from Gaussianity will likely be established by the detection of one or more of the brightest signals
using single source analysis techniques~\cite{Corbin:2010kt,Ellis:2012zv}.  The importance of their being few effective samples in the data is illustrated in
Figure 3, where correlation curves for various simulated signals are shown based on a ten year observation period (noise was not added to the signals in Figure 3 so as not to
obscure the intrinsic scatter from the Pulsar term, but a noise spectrum was used when computing the inner products). In these simulations the Pulsar timing
noise was assumed to have a white spectrum above 6 nHz, and a red spectrum at lower frequencies~\cite{Finn:2010ph}:
\begin{equation}
S_n(f) = {\rm const.}\left(1 + (f/6 \, {\rm nHz})^{-2} \right) \, .
\end{equation}

\begin{figure}[!ht]
  \centering
  \begin{tabular}{ccc}
  \hspace*{-4mm} \includegraphics[width=55mm]{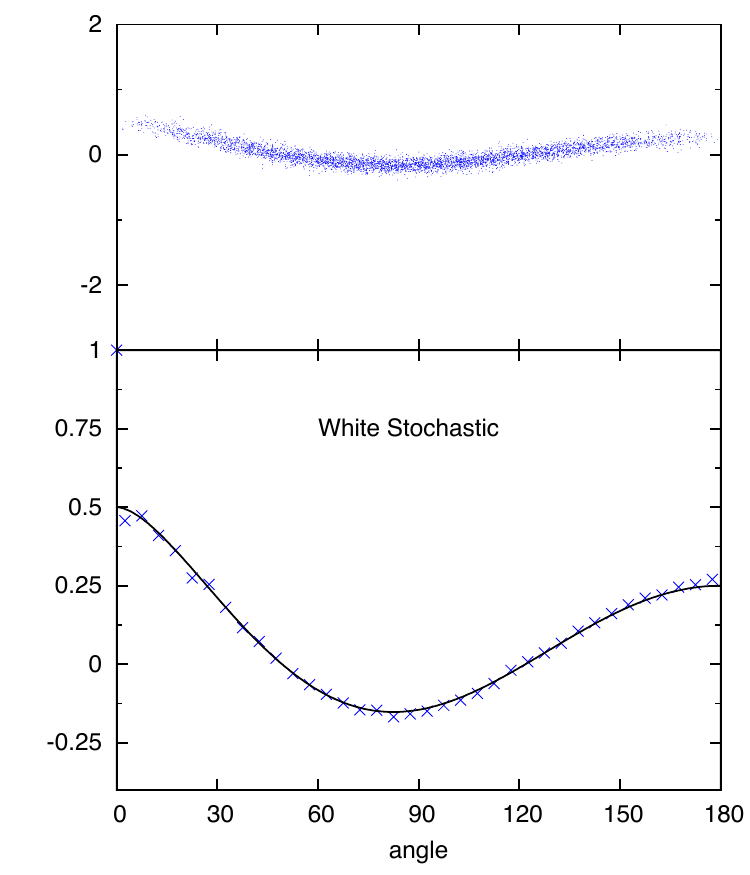}  &
   \hspace*{-8mm}  \includegraphics[width=55mm]{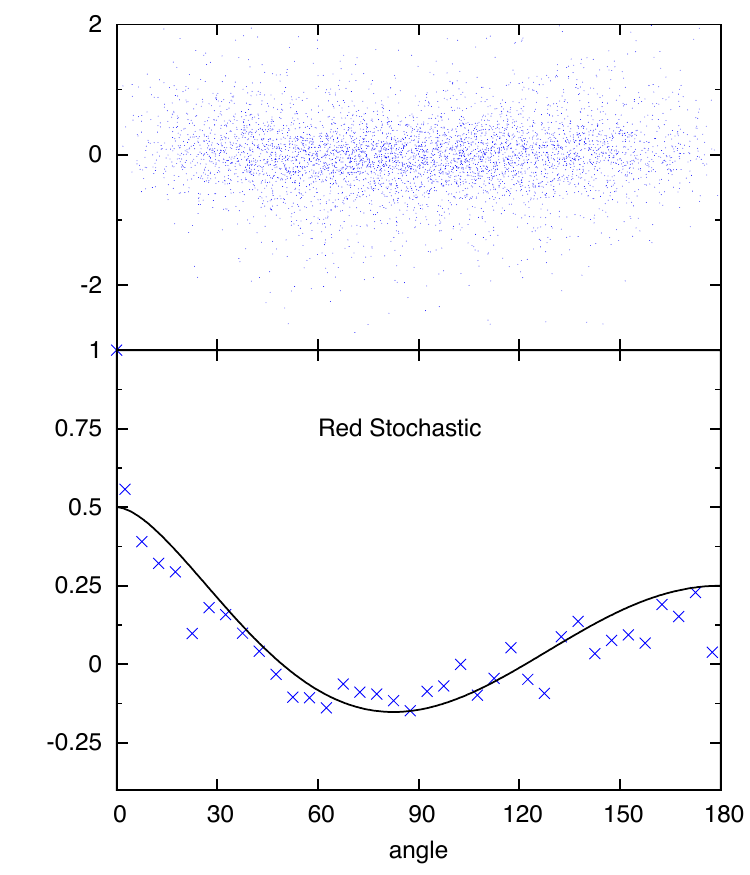}  &
     \hspace*{-8mm} \includegraphics[width=55mm]{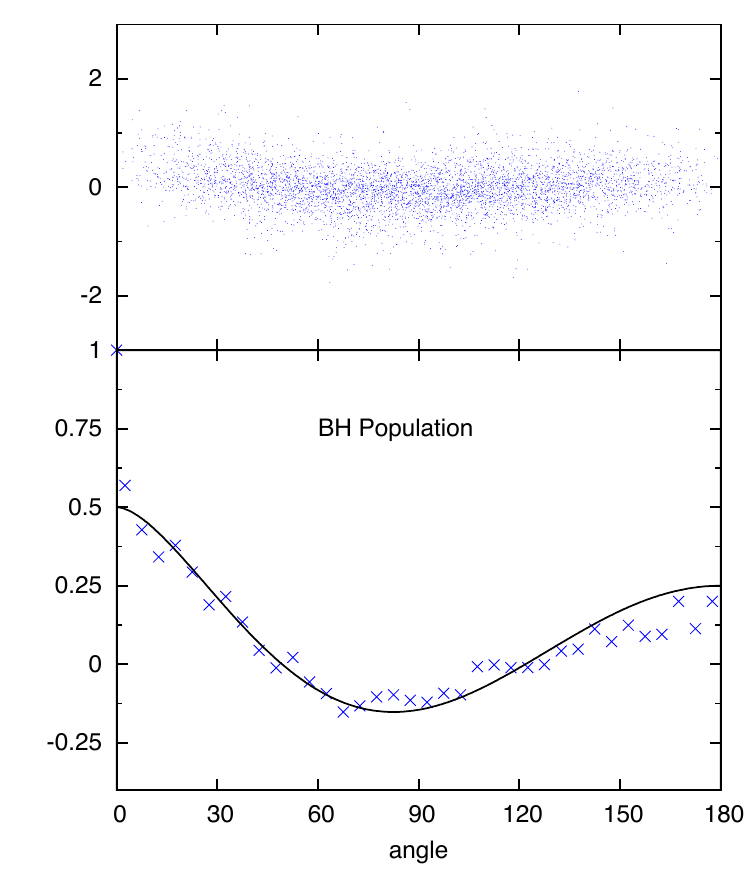}     \\
    \end{tabular}
  \caption{Simulated noise free correlation curves for (i) an isotropic, White Gaussian background using 100 frequency bins (ii) an isotropic Red Gaussian background
  with the spectrum predicted for a BH population (iii) a BH population model. The upper panels are raw scatter plots as a function of the angle between the Pulsars, while
  the lower panels average the correlations into 5 degree bins.}\label{fig:comp2}
\end{figure}

\noindent The first panel in Figure 3 shows the correlation curve for an isotropic stochastic background with a white spectrum that covers 100 frequency bins. The second
panel shows the correlation curve for an isotropic stochastic background with a red spectrum where the slope was chosen to match that from a population of
black hole binaries ($S_H(f) \sim f^{-13/3}$). The third panel shows the correlation curve for a realization of the BH population model used to generate Figure 1.
Remarkably, the scatter from the BH population is {\em less} than for an Gaussian stochastic background, as can be seen in the histograms of
$\langle r_i r_j \rangle/\langle H^2 \rangle-\alpha(\theta_{ij})$ shown in Figure 4.

\begin{figure}[!ht]
  \centering
  \includegraphics[width=100mm]{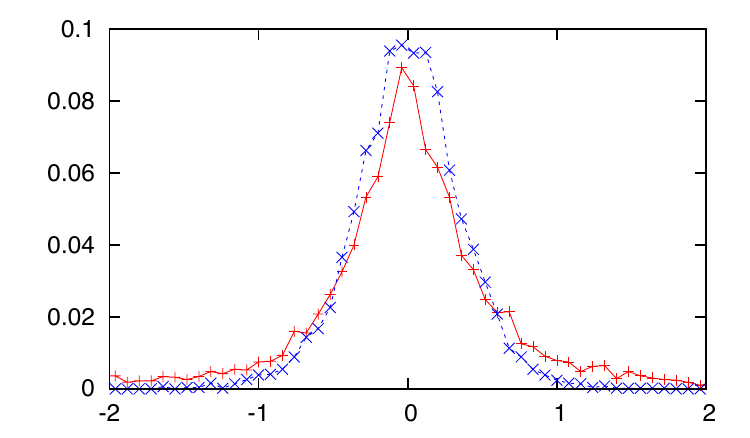}
  \caption{Histograms of the scatter in the correlation about the Hellings-Downs model for an isotropic, red Gaussian background (red, solid line), and for
  an astrophysical population of black hole binaries (blue, dashed line).}\label{fig:spread}
\end{figure}

Having established that the standard correlation analysis is capable of detecting the anisotropic, deterministic signal from an astrophysical population of black hole
binaries, it is worth considering how the analysis can be improved. What we are seeking is an analysis technique that has an optimal balance between fidelity in the
signal model and parsimony in terms of dimensionality. High dimensional models can achieve high fidelity, but at the cost of a larger trials factor (in a Frequentist setting)
or Occam factor (in a Bayesian setting). One high fidelity approach would be to abandon a correlation analysis in favor of a direct waveform template-based
search for individual systems~\cite{Corbin:2010kt,Ellis:2012zv}., along the lines of what has been proposed for detecting galactic binaries with a space based gravitational
wave detector~\cite{Crowder:2006eu}. The advantage of such an approach is that the signal model would accurately reflect the signals in the data, but the downside is
that it greatly increases the size of the parameter space to be explored. We may find ourselves in a regime where each individual source lies below the detection threshold,
while the combined signal may be detectable by some other less direct approach. A correlation analysis using a variant of (\ref{cross}), evaluated for several bright binaries
with particular orientations and sky location, and with the frequency domain inner products restricted to sub-bands where one or two signals dominate, may be effective,
but such an analysis introduces almost as many parameters as a multi-signal template based approach. One model that may find the sweet spot in the balance between
fidelity and complexity introduces a single orientation parameter per bright source which helps to account for the anisotropy of the underlying signal. The orientation parameter
is the angle between the actual signal direction and the signal direction used to construct the ``correlation template''. To see how this is derived consider the filtered
correlation function
\begin{equation}
\kappa_{ij}(\theta,\phi,\theta_T,\phi_T) = \langle r_i r_j \rangle(\theta,\phi)\,  \beta_{ij}(\theta_T,\phi_T) \, ,
\end{equation}
with
\begin{equation}\label{dfil}
\beta_{ij}(\theta_T,\phi_T) =  F_i^+(\theta_T,\phi_T)     F_j^+(\theta_T,\phi_T)  +  F_i^\times(\theta_T,\phi_T)  F_j^\times(\theta_T,\phi_T) \, .
\end{equation}
The filter $\beta_{ij}(\theta_T,\phi_T)$ is the polarization averaged correlation function for a point source at sky location $(\theta_T,\phi_T)$. Note that
sky average of this quantity is the standard Hellings-Downs correlation curve:
\begin{equation}\label{sba}
\frac{1}{4 \pi} \int \beta_{ij}(\theta_T,\phi_T)\, d\Omega_T=  \alpha(\theta_{ij}) \, .
\end{equation}
Averaging $\kappa_{ij}$ over Pulsar pairs separated by angle $\mu$ yields
\begin{equation}
\frac{1}{(4 \pi)^2} \,\int  \kappa_{ij}(\theta,\phi,\theta_T,\phi_T)\, \delta(\cos\mu - \hat{n}_i \cdot \hat{n}_j) \,  d\Omega_i d\Omega_j = \langle H^2 \rangle\,  \gamma(\mu,\zeta)
\end{equation}
where $\zeta$ is the angle between the source direction $(\theta,\phi)$ and the filter direction $(\theta_T,\phi_T)$. In the continuum limit, the standard $\rho$ statistic is recovered
by integrating the above expression over $\mu$ and $\zeta$: $\rho = \int \langle H^2 \rangle\,  \gamma(\mu,\zeta) \, d\cos\mu \, d\cos\zeta$. The function $\gamma(\mu,\zeta)$
is plotted in Figure 5. Note that the matched filter $\gamma(\mu,0)$ produces the largest correlation, and that using the sky averaged version of the filter ({\it i.e.} the average over $\zeta$) will degrade the sensitivity.  In practice, since the source location is {\it a priori} unknown, it is not possible to parametrize the
directional filter (\ref{dfil}) by the angle $\zeta$, and the search will have to be conducted using the parameters $(\theta_T,\phi_T)$. But despite the parameterization being two-dimensional, the physical search is still one dimensional since a circle of points on the $(\theta_T,\phi_T)$ sphere will yield exactly the same correlation curve $\gamma(\mu,\zeta)$. 

\begin{figure}[!ht]
  \centering
  \includegraphics[width=120mm]{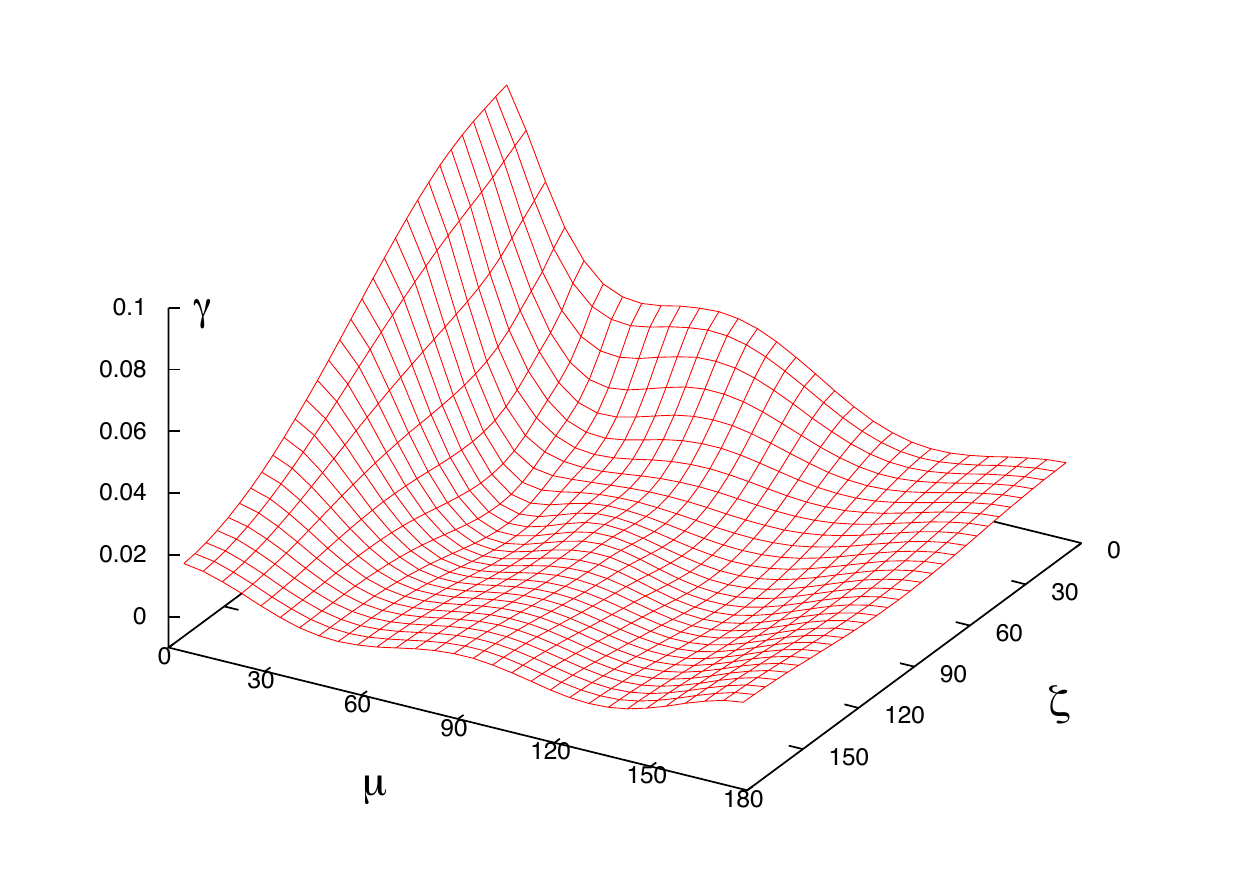}
  \caption{The directional correlation function $\gamma(\mu,\zeta)$.}\label{fig:gamma}
\end{figure}

For a black hole population the analysis could target the brightest black holes in each frequency band. For example, in the Bayesian formulation the correlation function
to be used in the likelihood (\ref{like}) could be generalized to
\begin{equation}
C_{ij}(f) = \sum_k S^k_H(f) \beta_{ij}(\theta^k_T,\phi^k_T) + S_{n_i}(f) \delta_{ij} \, ,
\end{equation}
where the $S^k_H(f)$ are localized to a particular frequency band. The optimal number of bands and their placement could be determined from the data using
transdimensional Markov Chain Monte Carlo techniques.

\section{Discussion}

We have shown that the standard cross-correlation analysis that was originally developed for isotropic, Gaussian backgrounds is capable of detecting the signals from
individual black hole binaries, and by extension, the combined signal generated by an astrophysical population of binaries. We have also argued that the standard
analysis will be sub-optimal in this case since the assumptions it makes about the signal are not valid, and we have suggested a number of approaches that may be
more sensitive. We are currently exploring the relative performance of the various methods using simulated data from a variety of population synthesis models, and
the results will be presented in a forthcoming publication.

\ack
This work was supported by NASA grant NNX10AH15G.


\Bibliography{99}

\bibitem{Sesana:2008mz} 
  A.~Sesana, A.~Vecchio and C.~N.~Colacino,
  arXiv:0804.4476 [astro-ph]
 
\bibitem{Sesana:2008xk} 
  A.~Sesana, A.~Vecchio and M.~Volonteri,
  arXiv:0809.3412 [astro-ph].
  
\bibitem{Sesana:2010mx} 
  A.~Sesana and A.~Vecchio,
  Class.\ Quant.\ Grav.\  {\bf 27}, 084016 (2010)
  [arXiv:1001.3161 [astro-ph.CO]].
\bibitem{Kocsis:2010xa} 
  B.~Kocsis and A.~Sesana,
  arXiv:1002.0584 [astro-ph.CO].
  
 \bibitem{CornishSesana}
 N.~J.~Cornish and A.~Sesana, 
 APS April Meeting Abstract APR.Q7005C, (2012) {\tt http://adsabs.harvard.edu/abs/2012APS..APR.Q7005C} .
 
 \bibitem{Ravi:2012bz} 
  V.~Ravi, J.~S.~B.~Wyithe, G.~Hobbs, R.~M.~Shannon, R.~N.~Manchester, D.~R.~B.~Yardley and M.~J.~Keith,
  Astrophys.\ J.\  {\bf 761}, 84 (2012)
  [arXiv:1210.3854 [astro-ph.CO]].
  
\bibitem{Hellings:1983fr} 
  R.~w.~Hellings and G.~s.~Downs,
  Astrophys.\ J.\  {\bf 265}, L39 (1983).

\bibitem{Jenet:2005pv} 
  F.~A.~Jenet, G.~B.~Hobbs, K.~J.~Lee and R.~N.~Manchester,
  Astrophys.\ J.\  {\bf 625}, L123 (2005)
  [astro-ph/0504458].

\bibitem{vanHaasteren:2008yh} 
  R.~van Haasteren, Y.~Levin, P.~McDonald and T.~Lu,
  arXiv:0809.0791 [astro-ph].
  
\bibitem{Bertone:2007sj} 
  S.~Bertone, G.~De Lucia and P.~A.~Thomas,
  Mon.\ Not.\ Roy.\ Astron.\ Soc.\  {\bf 379}, 1143 (2007)
  [astro-ph/0701407].
  
\bibitem{Springel:2005nw} 
  V.~Springel, S.~D.~M.~White, A.~Jenkins, C.~S.~Frenk, N.~Yoshida, L.~Gao, J.~Navarro and R.~Thacker {\it et al.},
  Nature {\bf 435}, 629 (2005)
  [astro-ph/0504097].
  
\bibitem{Gultekin:2009qn} 
  K.~Gultekin, D.~O.~Richstone, K.~Gebhardt, T.~R.~Lauer, S.~Tremaine, M.~C.~Aller, R.~Bender and A.~Dressler {\it et al.},
  Astrophys.\ J.\  {\bf 698}, 198 (2009)
  [arXiv:0903.4897 [astro-ph.GA]].

\bibitem{Finn:1998vp} 
  L.~S.~Finn,
  eConf C {\bf 9808031}, 07 (1998)
  [gr-qc/9903107].

\bibitem{Corbin:2010kt} 
  V.~Corbin and N.~J.~Cornish,
  arXiv:1008.1782 [astro-ph.HE].
\bibitem{Ellis:2012zv} 
  J.~A.~Ellis, X.~Siemens and J.~D.~E.~Creighton,
  Astrophys.\ J.\  {\bf 756}, 175 (2012)
  [arXiv:1204.4218 [astro-ph.IM]].
  
\bibitem{Finn:2010ph} 
  L.~S.~Finn and A.~N.~Lommen,
  Astrophys.\ J.\  {\bf 718}, 1400 (2010)
  [arXiv:1004.3499 [astro-ph.IM]].
  
\bibitem{Crowder:2006eu} 
  J.~Crowder and N.~Cornish,
  Phys.\ Rev.\ D {\bf 75}, 043008 (2007)
  [astro-ph/0611546].
  
\endbib
\end{document}